\documentclass{elsart}

\usepackage{natbib}

 \usepackage{graphicx}

\usepackage{amssymb}
\newcommand{\fr}{\frac}
\newcommand{\lf}{\left}
\newcommand{\rg}{\right}
\newcommand{\appleq}{\stackrel{<}{\sim}}
\begin{document}

\begin{frontmatter}



\title{Clausius' Virial vs. Total Potential Energy in the dynamics of a two-component system.}


\author{C. Marmo} \and{L. Secco}
\address{Department of Astronomy, University of Padova, Padova, Italy}
\ead{marmo@pd.astro.it}
\ead{secco@pd.astro.it}

\begin{abstract}
In a gravitational virialized
bound system built up of two components, one of which is embedded in the other,  
the Clausius' virial energy of one subcomponent is
not, in general, equal to its total potential energy, as 
occurs in a single system without external forces. This is the main reason
for the presence, in the case of two non-coinciding concentric spheroidal subsystems, 
of a minimum (in absolute value)
in the Clausius' virial of the inner component $B$,  when it assumes a special 
configuration characterized by a value of its semi-major axis we have named {\it tidal radius}.
The physical meaning, connected with its appearance, is to introduce a scale length on the
gravity field of the inner subsystem, which is induced from the outer one. Its
relevance in the galaxy dynamics  
has been stressed by demonstrating that some of the main features of Fundamental Plane may follow
as consequence of its existence. More physical insight into the dynamics of a two component
system may be got by looking
at the location of this scale length inside the plots of the potential energies of each
subsystem and of the whole system and by also taking into account the trend of the 
anti-symmetric residual-energy, that is the difference between the tidal and the interaction-energy
of each component. Some thermodynamical arguments related to the inner component are also added 
to prove as special is the {\it tidal radius configuration}. Moreover the role of the divergency at the center of the two subsystems in
obtaining this scale length is considered. For the sake of simplicity the analysis has been performed 
in the case of a frozen external component even if this constraint does not appear to be too relevant
in order to preserve the main results. 
\end{abstract}


\end{frontmatter}

\section{Introduction}
In order to understand the dynamics of a two-component system which a galaxy is, during its
virial evolution, we need to explain what the trends of the Clausius' Virial and of the total
potential energy of each subsystem, together with that of the whole system, are, as soon as the 
inner (baryonic) component
contracts inside the potential well of the outer dark matter halo.
In spite of the limits of this model which, of course,
is able to reproduce only some essential features of a real
galaxy, its relative analytical simplicity, due to the use of 
tensor virial
theorem, provides us with a powerful tool for understanding the 
role of the physical parameters involved in the dynamic 
evolution of a real system.
For the sake of simplicity we will
assume that the outer component is at fixed size and shape without considering this constraint
to be too essential in order to determine the main features of the dynamic
evolution we are dealing with ( see, \citep{S01} , hereafter LS1). Indeed
the masses of
the two components are not equal, the outer one being about ten times the inner one. As a
consequence, tidal 
influences between the subsystems are not symmetric; that acting from
inner to the outer
is actually
weaker than the reverse
(Caimmi \& Secco 1993).
Even if a contraction effect is
induced from inner density distribution on the inner regions of
outer halo, during the dynamical evolution, as already underlined by 
Barnes \& White (1984) , that effect  
seems, indeed, does not cause  
a dramatic modification of the outer mass distribution, if there is 
a supernovae- driven outflow, according to 
some 
recent N-body simulations (Lia 
et al. 2000). On the other side, models in which this constraint has been changed with
some less stringent additional conditions (Caimmi 1994), seem to prove the same conclusion.
Then, in order to underline the main 
effect of inducing
a tidal scale length from the dark halo over the luminous component, we will
neglect 
this possible modification.
It seems to us that this may correspond to change from a given 
mass distribution to
an other with a new exponent $d$ (see, sect.3) not too different from the previous
one; that might be included in the different cases considered
without cancelling the effect we are dealing with.

 The two subsystems will be modelled by two penetrating similar-strata spheroids with two power-law 
density profiles. The advantage of this model is to be able
 to handle all the energetics of a two-component system during its dynamic evolution in an 
 analytical way. Nevertheless to obtain some 
 physical meaningful relationships, we prefer to 
 sacrifice the more correct, but very  boring, exact analytical expressions and 
 to use some suitable mathematical approximations. Through them it will be possible to describe the whole 
 virial evolution by means of relatively simple equations which involve only the 
 two exponents ($b$ and $d$ of sect.\ref{SSSS}) which characterize 
 the two subsystem power-law density 
 distributions ($B$, the inner and $D$ the outer, respectively), as soon as the common form 
 factor $F$  of sect.\ref{SSSS} is fixed (=2, in the spherical case).
 The appearance of a non-monotonic trend for the inner Clausius' virial energy, the contrary of that
 which refers to the inner total potential energy, is the most important result for the dynamics of a
 two-component system during its virial phase. Its relevance  
has already been stressed by demonstrating how some of the main features of elliptical 
galaxy Fundamental Plane may follow
as a consequence of it (LS1). Its occurence seems to be a general feature when one component is completely
embedded in another. This has also been tested by considering different models from those taken
into account here as such the case of heterogeneous density
 profiles with different eccentricities \citep{CS01}. 
 
 Moreover, how much the extensions of the cores, at the center of the two substructures, have to be in 
 order to obtain 
 a maximum of the virial energy in the inner component, will be investigated when the luminous matter
 distribution becomes particularly steep.
The present scheme may be extended to other two-component systems built up of baryonic + dark matter 
as, e.g., the galaxy clusters are.

\section{Virial and total potential energy}
\label{VTPE}
By referring to Caimmi \& Secco (1992) and the references therein, the Clausius' virial, or 
virial potential energy, $V_u$ ($u=B,D$), appears with the kinetic energy, $T_u$, in
the scalar virial equations related to the two components, the inner, $B$, embedded in the outer, $D$, of a
virialized system:
\begin{equation}
\label{fun}
2T_u+V_u=0.
\end{equation}
It refers to the energy contribution due to the two active forces on the respective subsystem:
the self-gravity and the tidal-gravity due to the other component. 

Generally speaking, it turns out to be:
\begin{equation}
V_u=\Omega_u+V_{uv},
\label{Vu}
\end{equation}
where,
\begin{equation}
\Omega_u=\int\rho_u\sum_{r=1}^3x_r\frac{\partial\Phi_u}{\partial x_r}d\vec{x}_u,
\end{equation}
and
\begin{equation}
(V_{uv})=\int\rho_u\sum_{r=1}^3x_r\frac{\partial\Phi_v}{\partial x_r}d\vec{x}_u,
\end{equation}
$\Phi_u$ and $\Phi_v$ being the gravitational potentials due to $u$-matter an $v$-matter distributions, respectively.

In order to obtain
the total potential energy of the $u$-component, we have to add at the self-potential energy
the interaction energy $W_{uv}$, i.e.:
\begin{equation}
\label{eq5}
(E_{pot})_u=\Omega_u+W_{uv},
\end{equation}
The difference between the total and the virial $u$-energy is:
\begin{equation}
\label{eq6}
V_u-(E_{pot})_u=Q_{uv},
\end{equation}
$Q_{uv}$ being the residual energy, which is anti-symmetric   
with respect to the exchange of one component with the other. 
The following equations also hold:
\begin{eqnarray}
\label{scar1}
V_{uv} &=& W_{uv}+Q_{uv},\\
W_{uv} &=& W_{vu};
\end{eqnarray}
the latter shows the symmetry of the interaction energy tensor.

As consequence,
the total potential energy tensor trace of the whole system, in the principal inertia axes frame of
reference, is expressed by:
\begin{equation}
\label{eq7}
E_{pt}=(E_{pot})_u+(E_{pot})_v=\Omega_u+\Omega_v+W_{uv}+W_{vu}=V_u+V_v.
\end{equation}

It should be underlined that the residual energy may become zero, and then the total potential
energy of each subsystem is equal to the Clausius' virial energy, only for some specific 
configurations which depend on the mass distributions of the two subsystems (see,
sect.\ref{DM}).

\section{Similar and similar-strata spheroids}
\label{SSSS}
According to \citet{C93} (hereafter RC) and to \citet{S00} (hereafter LS), we consider, now, the case of two similar spheroids
(that means, with the same axis ratio $\epsilon_B=\epsilon_D=\epsilon$ (then the same form factor
$F=(2\frac{\alpha (\epsilon)}{\epsilon}+\frac{\epsilon^2\gamma (\epsilon )}{\epsilon})$), with similar and coaxial strata mass
distributions. We set the radial density profiles according to two power-laws as follows:
\begin{eqnarray}
\rho_B &\sim& \frac{1}{r^b}, \\
\rho_D &\sim& \frac{1}{r^d};
\end{eqnarray}
and to avoid the central divergence we add an inner homogeneous core to both mass distributions.
For the sake of simplicity we assume here that these two cores have the same extension ($\xi_c$) in the
a-dimensional coordinates (see, RC).
Then, the traces of B and D self-potential energy tensors are, respectively:
\begin{eqnarray}
\Omega_B &=& -\nu_{\Omega B}\frac{GM_B^2}{a_B}F, 
\label{OB}\\
\Omega_D &=& -\nu_{\Omega D}\frac{GM_D^2}{a_D}F.
\label{OD}
\end{eqnarray}
where $M_u$  is the mass and $a_u$ the semi-major axis of the considered subsystem. 
The coefficient $\nu_{\Omega u}$ depends on the mass distribution via the functions $h$ and $g_{3u}$, from
Table 1 of the Appendix A in RC.

The traces of B and D tidal-potential energy tensors are:
\begin{eqnarray}
\label{vbd}
V_{BD} &=& -\nu_V\frac{GM_B^2}{a_B}F, 
\label{VBD}\\
V_{DB} &=& -\nu_{VD}\frac{GM_D^2}{a_D}F.
\label{VDB}
\end{eqnarray}
The symmetric interaction energy tensor trace is
\begin{equation}
\label{wbd}
W_{BD}=-\nu_W\frac{GM_B^2}{a_B}F.
\end{equation}
The following expressions define the $\nu$-coefficients of the previous energies:
\begin{eqnarray}
\label{nuvi}
\nu_V &=& -\fr{9}{8}\lf[(\nu_B)_{M}(\nu_D)_{M}\rg]^{-1}mw^{ext}(x), \\
\label{nuvd}
\nu_{VD} &=& -\fr{9}{8}\lf[(\nu_B)_{M}(\nu_D)_{M}\rg]^{-1}\fr{1}{mx}w^{int}(x), \\
\label{nuw}
\nu_W &=& -\fr{9}{16}\lf[(\nu_B)_{M}(\nu_D)_{M}\rg]^{-1}m[w^{int}(x)+w^{ext}(x)],
\end{eqnarray}
where $x=a_B/a_D$ e $m=M_D/M_B$. The  $(\nu_u)_{M}$ coefficients are defined in Tab. 1 of RC.
Due to the eqs.(\ref{nuvi}, \ref{nuw}) the following relation also holds:
\begin{equation}
\label{rel}
W_{BD}=\frac{9}{16}[(\nu_B)_M(\nu_D)_M]^{-1} 
 m~w^{int}(x)\frac{GM^2_B}{a_B}F+\frac{1}{2}V_{BD}
\end{equation}
The definitions of $w^{int}(x)$ and $w^{ext}(x)$ and their approximations are given in 
the Appendix.

It should be stressed that $w^{ext}$ contains the expression of the $D$-mass fraction which has 
dynamical effect over the 
$B$-component at every value of $x$ (see Appendix).

\section{Dynamical quantities}
In order to gain some physical meaningful relationships, we will adopt here some mathematical
simplifications.
It should be underlined that all the approximations
we consider in the following section and in the Appendix, refer 
to the case in which the two cores have the same a-dimensional extension $\xi_c$ and under the 
limitation that $x > \xi_c$. In a further paper 
we will relax these constraints.

Under the approximations we have considered in the Appendix, eq.(\ref{vbd}) becomes:
\begin{equation}
\label{Avbd}
V_{BD}\simeq- \nu'_V~G \frac{M_B \widetilde{M_D}}{a_B}F
\end{equation}
$B$-Clausius' virial may be defined as:
\begin{equation}
\label{Cla}
 V_B \simeq GM_B^2F\Big[-\nu_{\Omega B}\frac{1}{a_B}- \nu'_V 
		  \frac{M_D}{M_B}\frac{1}{a_D} (\frac{a_B}
           {a_D})^{2-d}\Big]
\end{equation}
or in normalized form:
\begin{equation}
\label{virn}
\widetilde{V_B}=\frac{V_B~ a_D}{GM_B^2F }\simeq-\frac{\nu_{\Omega B}}
		   {x}-\nu'_V\Big(\frac{M_D}{M_B}\Big) x^{2-d};~~x=\frac{a_B}{a_D}
\end{equation}

By using the eq.(\ref{rel}) and the approximation (\ref{Approx})
($\epsilon\simeq 0$) we obtain:
\begin{equation}
\label{wapprox}
W_{BD}\simeq -\frac{9}{4}[(\nu_B)_M(\nu_D)_M]^{-1}\frac{GM_BM_D}{a_D} 
 \frac{\xi_c^{b+d}}{(2-d)(3-b)}F-\frac{1}{2(2-d)}V_{BD}
\end{equation}
and from eq.(\ref{scar1}):
\begin{equation}
\label{qapprox}
Q_{BD}\simeq \frac{9}{4}[(\nu_B)_M(\nu_D)_M]^{-1}\frac{GM_BM_D}{a_D} 
 \frac{\xi_c^{b+d}}{(2-d)(3-b)}F+\frac{5-2d}{2(2-d)}V_{BD}
\end{equation}
Finally from the definition of $(E_{pot})_B$ (eq.(\ref{eq5})) and by the 
approximations of eqs.(\ref{wapprox}, \ref{Avbd}), we have:
\begin{eqnarray}
\label{Atotpot}
(E_{pot})_B&\simeq& \Omega_B
-\frac{9}{4}[(\nu_B)_M(\nu_D)_M]^{-1}\frac{GM_BM_D}{a_D} 
 \frac{\xi_c^{b+d}}{(2-d)(3-b)}F
 \nonumber\\
 & &
 + \frac{\nu'_V}{2(2-d)}~G \frac{M_B^2 m }{a_B}
 (\frac{a_B}{a_D})^{3-d}F
\end{eqnarray}
By normalization to the same factor of $\widetilde{V_B}$ (eq.(\ref{virn})), it
follows:
\begin{equation}
\label{ABpot}
(\widetilde{E}_{pot})_B\simeq -\frac{\nu_{\Omega_B}}{x}
-\frac{9}{4}[(\nu_B)_M(\nu_D)_M]^{-1}m 
 \frac{\xi_c^{b+d}}{(2-d)(3-b)}+ \frac{\nu'_V}{2(2-d)}~m 
 x^{2-d}
\end{equation}
or:
\begin{equation}
\label{Atopot1}
(\widetilde{E}_{pot})_B\simeq -\frac{\nu_{\Omega_B}}{x}
-\frac{1}{2}\nu'_Vm 
 \frac{(3-d)[5-(b+d)]}{(2-d)(3-b)}+ \frac{\nu'_V}{2(2-d)}~m 
 x^{2-d}
\end{equation}
where is:
\begin{equation}
\label{wnorm}
\widetilde{W}_{BD}=\widetilde{W}_{DB}\simeq
-\frac{1}{2}\nu'_Vm 
 \frac{(3-d)[5-(b+d)]}{(2-d)(3-b)}+ \frac{\nu'_V}{2(2-d)}~m 
 x^{2-d}
\end{equation}
Moreover the residual energy of eq.(\ref{qapprox}) becomes:
\begin{equation}
\label{qnorm}
\widetilde{Q}_{BD}=-\widetilde{Q}_{DB}\simeq
\frac{1}{2}\nu'_V~m 
 \frac{(3-d)[5-(b+d)]}{(2-d)(3-b)}- \frac{(5-2d)}{2(2-d)}\nu'_V~m 
 x^{2-d}
\end{equation}

The Clausius' virial related to the $D$ subsystem, eq.(\ref{Vu}), is:
$$V_D=\Omega_D+V_{DB}$$
where $\Omega_D$ and $V_{DB}$ correspond to the eqs.(\ref{OD}, \ref{VDB}),
respectively.
In the limitations under which the approximation (\ref{nuaprx}) holds, 
the approximative tensor trace of Clausius' virial, related to
$D$ component, becomes:
\begin{equation}
\label{vdd}
V_D\simeq-\nu_{\Omega D}G\frac{M_D^2}{a_D}F-\nu'_VG\frac{M_DM_B}{a_D}F\Big\{\frac{(3-d)[
5-(b+d)]}{(2-d)(3-b)}-\frac{3-d}{2-d}\Big(\frac{a_B}{a_D}\Big)^{2-d}\Big\}
\end{equation}
By normalization at the usual factor $\frac{GM^2_BF}{a_D}$, we obtain:
\begin{equation}
\label{vddn}
\widetilde {V_D}\simeq-\nu_{\Omega D}m^2-\nu'_Vm\frac{3-d}{2-d}\Big\{\frac{[
5-(b+d)]}{(3-b)}-x^{2-d}\Big\}
\end{equation}
By using the definition of the total potential energy of $D$-component
of eq.(\ref{eq5}) and passing through the eq.(\ref{wnorm}), we obtain:
\begin{equation}
\label{epotdn}
(\widetilde{E}_{pot})_D\simeq -\nu_{\Omega D}m^2-\frac{1}{2}\nu'_Vm 
 \frac{(3-d)[5-(b+d)]}{(2-d)(3-b)}+ \frac{\nu'_V}{2(2-d)}~m 
 x^{2-d}
\end{equation}

The definition of total potential energy of the whole system, eq.(\ref{eq7}),
yields the following normalized and approximate expression:
\begin{equation}
\label{eptnor}
\widetilde{E}_{pt}\simeq-\frac{\nu_{\Omega B}}
		   {x}-\nu_{\Omega D}m^2-\nu'_Vm 
 \frac{(3-d)[5-(b+d)]}{(2-d)(3-b)}+ \frac{\nu'_V}{(2-d)}~m 
 x^{2-d}
\end{equation}

As a concluding remark of this section we wish open the question of which is the weight of the 
approximations
on the trends of the energies if they are compared with those coming from the rigorous analytical
expressions. This is remarkable (see the next section) when it refers to the non-monotonic character
of the luminous Clausius' virial energy  which turns out to be of extremle
importance for
the dynamics of the 
two-component system. Generally speaking, the approximations 
considered ($b<2$) are as good as $\xi_c \rightarrow 0$ (e.g., the maximum score for the main 
physical quantities
of Tab.1 is of $0.5\%$ if $\xi_c=0.01$ and $b=1.5$), and that will constraint us to explore
what happens on this limit.

\section{Luminous component energy trends}
\label{BM}
We will analyze 
the trends of the two energies, the Clausius' virial of eq.(\ref{Vu}) and the total potential 
energy of eq.(\ref{eq5})
normalized by the factor $\frac{FGM^2_B}{a_D}$, which correspond
to the $B$ component mass distribution (here assumed to be the
luminous one). They are approximated by the eq.(\ref{virn}) and
the eq.(\ref{Atopot1}) respectively. The related rigorous functions are:
\begin{equation}
\widetilde{V}_B=-\frac{\nu_{\Omega B}}{x}-\frac{\nu_V}{x}
\end{equation}
and:
\begin{equation}
(\widetilde{E}_{pot})_ B=-\frac{\nu_{\Omega B}}{x}-\frac{\nu_W}{x}
\end{equation}
Their corresponding trends are given in the Fig.1 for the cases 2) and 4) of Table 1, by 
assuming $\xi_c=0.01$ and $m=8.5$. From the comparison between the values of the main physical quantities
of Tab.1, which are evaluated by using the rigorous trends shown in the figures,
and those
yielded by the approximate formulae, the conclusion is that the errors are less than $0.5\%$. This maximum
error 
is reached only in the case 4).
\begin{figure}[!h]
\begin{center}
\includegraphics[height=8cm,width=12cm]{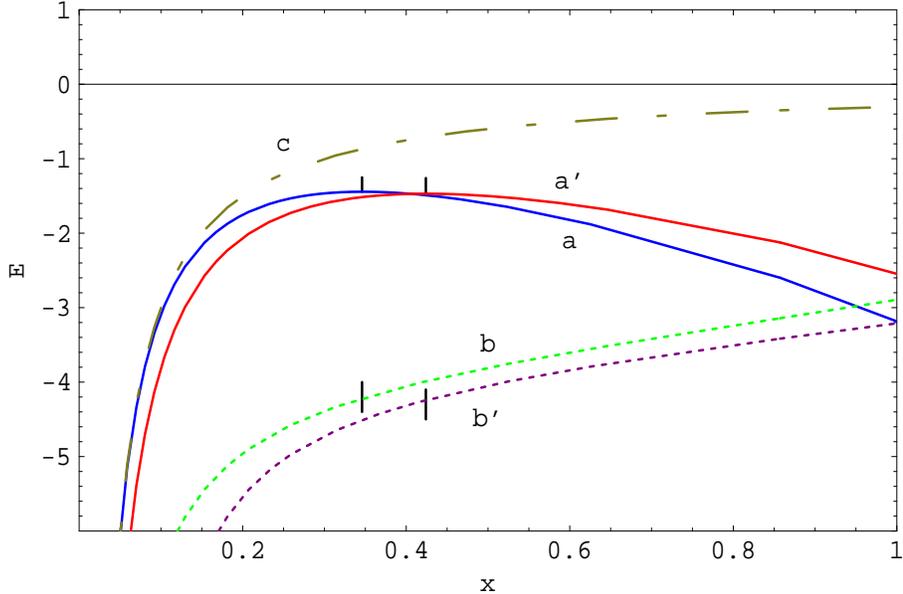}
\caption{The energy trends of the $B$-system as function 
of $x=a_B/a_D$, normalized at the factor $(GM_B^2F)/a_D$. The a-curve 
(solid line) represents the Clausius' virial 
energy, $\widetilde {V}_B$, and the b-one (dotted line), the total potential 
energy, $\widetilde {E}_B$, (case 2) of Tab.1); a' and b' are the corresponding curves
for the case 4) of Tab.1.
For comparison the c-curve (long short-dashed line) is the self-energy trend of a single 
$B$-component in the case 2).
 The vertical marks correspond to the {\it tidal radius} configurations.}
\label{marmo1}
\end{center}
\end{figure}

It appears the relevant non-monotonic character of the Clausius' virial energy (curves a, a') 
in respect to the
monotonic one of the total potential energy of the same component (curves b, b'). 

As already shown
(see, LS and LS1),
when we describe a gravitational structure as in section \ref{SSSS}, a configuration 
which minimizes the
Clausius' virial of inner component exists, if the outer density distribution 
is described by a power law with an exponent less then $2$.
This special configuration of the embedded component is also analytically defined by the d'Alembert's Principle 
of virtual works. The condition of virial equilibrium
means that, if the virial energy has a 
minimum (in absolute value), the total kinetic energy also exhibits a 
minimum. This implies, in turn, that if the kinetic energy of the inner bright component is
dominated by the random 
velocity dispersion
(as in ellipticals) the corresponding
macroscopic pressure support has a minimum; if, on the contrary, it is the rotational 
kinetic energy which dominates (as in the spirals), the rotation velocity distribution has to be
that which is able to minimize the corresponding total ordered kinetic energy.
The semi-major axis $a_t$, which characterizes this
configuration, is a scale length acquired by the gravitational field of the 
inner component, typically without any. It is expressed, by deriving the approximated
eq.(\ref{virn}) in respect to $x$:
\begin{equation}
a_t=\lf(\fr{\nu_{\Omega B}}{\nu'_V}\fr{1}{(2-d)}\fr{M_B}{M_D}\rg)^{\fr{1}{3-d}}a_D,
\label{at}
\end{equation}
The result is the same as \citet{H58} found for the tidal radius of
a spherical star cluster 
embedded in the galaxy tidal potential and represents
a generalization of it. This is the reason the dimension given by 
the eq.(\ref{at}) is named as
{\it{tidal radius}}.
By assuming $a_D=const.$, $M_D/M_B\simeq 10$ (on the basis of cosmological arguments) and
by starting with the dynamical evolution at $a_B=a_{Bo}=a_D$ ($x$=1), we can see from eq.(\ref{virn})
that the ratio between the tidal-energy term ($\widetilde {V}_{BD}$, the second one) and the self-energy
term ($\widetilde {\Omega}_B$, the first one) is:
\begin{equation}
\label{zeta}
\zeta_1=\Big(\frac{\widetilde{V}_{BD}}{\widetilde{\Omega}_B}\Big)_{x=1}\simeq\frac{\nu'_V}{\nu_{\Omega B}}m
\end{equation}
For all the realistic values of the couple ($b$,$d$) we have considered (see, Tab.1 and Fig.1 in LS)
the ratio $\zeta_1$ is greater than one and 
then the tidal-energy term is
dominant with respect to the self-energy one. As $a_B$ decreases at $a_D=const.$, the 
tidal-energy decreases, in absolute value,
because of the decrease in $D$-mass inside 
the $B$-boundary, which is $M_Dx^{3-d}$.
Then the integral of force times the the position, $V_{BD}$, (\ref{Avbd}) decreases 
as $x^{2-d}$,
meanwhile the self-energy increases in absolute value as $x^{-1}$. The balance 
between the self- and tidal-energy is reached
at about $a_t$ and then, when $x$ becomes less than $x_t$ ($=a_t/a_D)$, the 
self-energy
overwhelms the tidal-energy.
The $\zeta_t$ ratio between the two energies at this minimum becomes:
\begin{equation}
\zeta_t\simeq\frac{1}{2-d}
\end{equation}
The approximated eq.(\ref{virn}) allows us to connect analytically the 
$x_e$ value at which the two energies, self and tidal, are equal with  
$x_t$.  Indeed, it turns to be:
\begin{equation}
\label{xe}
x_e\simeq x_t(2-d)^{\frac{1}{3-d}}
\end{equation}
If $d=1$ the equipartition between the two energies properly occurs at $x_t$.
 
Neverthless the equation (\ref{at}) is not the most general expression: the two 
components 
considered here are, indeed, always
homothetic as $a_B$ changes. In the homogeneous ($b=d=0$), non similar 
case, when the axis ratio 
depends weakly on $a_B$,
one approximative analytical expression of $a_t$ exists (see, eq.(53) in LS).
In order to look for a more general extension, as in the case of two
different eccentricities with two heterogeneous density profiles, only a first 
order approach has been taken into account \citep{CS01}, because in this case the  
tensor, $(V_{uv})_{ij}~ (u,v=B,D;~i,j=x,y,z)$, in an analytical
form, is not yet available.

It is interesting to consider where the tidal radius is located on the energies curves , in order to
gain more insight into its physical meaning.
$x_t$ (the vertical mark on curves of Fig.1,) corresponds to the beginning of 
the self-energy domain in the Clausius' energy trend. Indeed, as soon as $~x<~x_t$  
both the virial and the total potential energy of the luminous component begin to 
increase (in absolute value) very steeply
as the size decreases, with approximately the same trend of a single $B$-component
self-energy (curve c in Fig.1).
At dimension greater than $x_t$,
the total potential 
energy curve of $B$-subsystem exibits slopes smaller than the previous ones at 
$x<x_t$ (see curve b of Fig.2), by changing
its concavity a little bit before $x=1$ (the zero of the c-curve(= $\frac{d^2(\widetilde {E}_{
pot})_B}{dx^2}$) in Fig.2). If one looks at the derivative $\frac{d(\widetilde {E}_{
pot})_B}{dx}$ , curve b of Fig.2, it is manifest that $a_t$ gives the scale length 
at which the slope of the $B$ total potential energy changes. Indeed, if for $x~<~x_t$ 
the slope is about the same of that of a single contracting component (curve b'), as soon as $x~>~x_t$ 
the slope changes step by step in order to stabilize itself on 
one, which characterizes the tidal-energy
domain, at about $x\simeq 3 x_t\simeq 1$. It should be noted that the flex point is 
included  inside the
interval $x_t-3x_t$ (see Tab.1, even if in the case $4)$, strictly speaking, the mathematical 
flex point
does not correspond to a physical configuration in our model because its coordinate is a little bit
greater than $1$). For $x>x_t$ the rate at which the
slope of the $B$ total potential energy (curve c) varies is smaller than 
that at which the slope 
of same component
without the potential well of dark matter (curve c') varies.

\begin{figure}[!h]
\begin{center}
\includegraphics[height=8cm,width=12cm]{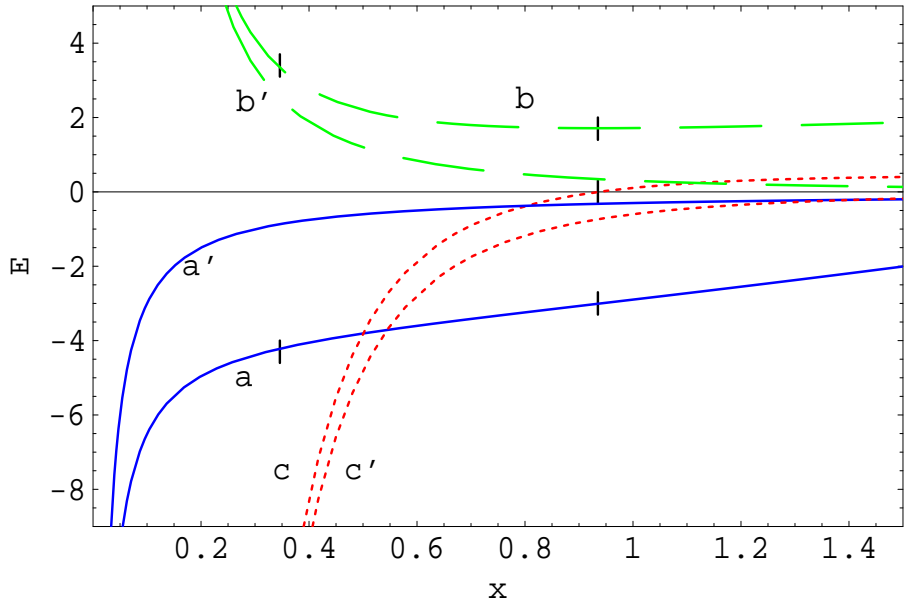}
\caption{Detailed potential energy trend for the $B$-system with and without the dark
potential well, as function of $x$. The normalization factor
is the same as in Fig.1. The a-curve (solid line) represents the total potential 
energy, $(\widetilde {E}_{pot})_B$ in the case 2) of Tab.1; the b (long dashed line) and 
c (dotted line) represent 
the first and the second derivative of the total potential energy, respectively. b' and c'
are the same derivatives for the potential energy of the single $B$ component.
The vertical marks on the left correspond to the {\it tidal radius} configuration,
the other, on the right, to the {\it flex} point of coordinate $x_f$ (see, Tab.1).}
\label{marmo2}
\end{center}
\end{figure}

In order to understand under which conditions the flex point does appear and where it is located, we
will take into account the approximation given by eq.(\ref{Atopot1}). Indeed, it
allows us to handle the following analytical expressions.
At fixed outer component, the first and second derivative of $(\widetilde{E}_{pot})_B$
are, respectively:
\begin{eqnarray}
\label{first}
\frac{d(\widetilde {E}_{
pot})_B}{dx}\simeq \frac{\nu_{\Omega_B}}{x^2}
+\frac{1}{2}\nu'_Vm  
 x^{1-d} \\
\label{second}
\frac{d^2(\widetilde {E}_{
pot})_B}{dx^2}\simeq -2\frac{\nu_{\Omega_B}}{x^3}
+\frac{1}{2}\nu'_Vm  
 \frac{1}{x^{d}}(1-d)
\end{eqnarray}
Then the flex point will occur at:
\begin{equation}
\label{flex}
x_f\simeq \big(4\frac{2-d}{1-d}\big)^{\frac{1}{3-d}}x_t
\end{equation}

It should be noted that the flex point may exist only if $d<1$ and
that it occurs at $x_t$ times a factor $f=\big(4\frac{2-d}{1-d}\big)^{\frac{1}{3-d}}$ (see, Tab.1) 
which depends only on
the exponent $d$ and which changes in a monotonic way from $2$ to
$\infty$ as $d$ increases from $0$ to $1$.
\begin{table}
\caption{The most important physical parameters in the energy trends of figures. The common curve
parameters are: $\xi_c=0.01,~m=8.5$;
$x_t=a_t/a_D$; $x_f$ marks the ratio $a_B/a_D$ at which the variation of total potential energy
of the $B$-component reaches its minimum (the flex, eq.\ref{flex}); $x'_f$ marks the flex coordinate 
for the total potential energy
of the whole system (eq.\ref{tflex}). $f=x_f/x_t$.}
\vspace{0.3cm}
\begin{center}
\footnotesize
\begin{tabular}{cccccccc}
\hline
&&&&&&&\\
cases &$b$&$d$&$x_t$&$f$&$x_f$&$3x_t$&$x'_f$\\
&&&&&&&\\
\hline
&&&&&&&\\
1)&$0.0$&$0.0$&$0.389$&$2.000$&$0.778$&$1.167$&$0.590$\\

2)&$0.0$&$0.5$&$0.346$&$2.702$&$0.935$&$1.039$&$0.708$\\
3)&$0.5$&$0.5$&$0.361$&$2.702$&$0.975$&$1.083$&$0.739$\\
4)&$1.5$&$0.5$&$0.425$&$2.703$&$1.149$&$1.274$&$0.871$\\
&&&&&&&\\
\hline
&&&&&&&\\

\hline

\end{tabular}
\end{center}
\label{tab1B}
\end{table}

  We now turn back on the question, opened at the end of the previous section, related to the 
 dependence of the non-monotonic character
of the luminous Clausius' virial energy, and of the location of its maximum, on the
approximations used. As we have seen the approximated expressions of energies 
tend to the rigorous ones as soon as the $\xi_c \rightarrow 0$. Then    
we have to explore
what happens when the core dimensions in the two components decrease. 
In accordance with our previous discussion on this point (see, sect.A.2 in LS),
we again conclude that, if the exponent $b$ is not 
too high ($b<2$) and the two power-law exponents satisfy the constraints given 
by (\ref{limits}), the maximum
on the $V_B$ trend appears and its coordinate $x_t$ changes very little as $\xi_c\rightarrow 0$.
As soon as $2\appleq b <~2.5$ the maximum also exists under the limitations (\ref{limits}), but
its location moves towards $x=1$ with $\xi_c\rightarrow 0$, by reaching a limit coordinate
which is always less than $1$.
When $2.5\le b ~<~3$ and again the limitations (\ref{limits}) are satisfied, the maximum 
appears if $\xi_c$ is not too small. Its coordinate $x_t$ moves towards $1$ as soon as 
$\xi_c$ decreases and then it disappears as $\xi_c\rightarrow 0$ (in the case of, b=2.5, 
which happens at $\xi_c=0.00038)$. These last two results are something
new to add to the previous analysis.
Then the following warning has to be pointed out:  
the non-monotonic character of $V_B$, may be lost\footnote{It may be conserved from a mathemathical point of 
view, but the maximum corresponds to a non-physical configuration in our model, because it occurs at $x_t>1$.} if the dimensions 
of the homogeneous cores become
very small with $2.5\le b$, even if the conditions (\ref{limits}) are fulfilled. To 
have or not to have the divergency at the center of the two configurations, seems
to become an essential feature in order to obtain or not an induced scale length on the inner component. 
This new message 
again requires, with the observations made at the beginning of the previous section, a further 
paper on this matter.

\section{Thermodynamical arguments}

To underline as special is the {\it tidal radius configuration} we will consider here some
aspects of the thermodynamics related to the luminous component. As well kwown (Chandrasekhar 1939)
when a single virialized component evolves through a quasi-static sequence of contractions, it has to 
lose an heat amount, $\Delta Q$, which is one half of the variation of its self potential 
energy due to the contraction, in the meanwhile the other half goes to
heat the structure. That means we need of a dissipation mechanism (e.g., gas cloud collisions) and
of cooling processes. The consequence will be a local decreasing of the structure entropy and a corresponding
increase of the universe entropy according to the $II^o$ thermodynamic principle (Secco 1999).
Now, the question is: what happens if we take into account the contraction process of the 
same component ($B$) if it is embedded
inside an other non-dissipative one ($D$)? In order to answer the question we have to re-consider 
the two fundamental equations:
1) the first thermodynamic Principle applied to $B$ system and ii) the variation of the virial 
quantities 
of the same component during a
quasi-static transition. 
In the $I^o$ Principle equation it has to appear now the work done by both the forces on the 
system: the self 
gravity and the gravity which the dark matter distribution exerts on it. The corresponding 
potential energy
variation for a small contraction will be $\Delta V_B$. 
Combining the two equations and considering that the 
internal energy
is the total kinetic energy of the system, $T_B$, (Schwarzshild 1958), we obtain:

\begin{eqnarray}
\label{varvir1}
\Delta Q=\Delta V_B/2\\
\label{varvir2}
\Delta T_B=-\Delta V_B/2
\end{eqnarray}
where $\Delta Q$ is the amount of heat exchanged between the system and the universe.
Again the two requests that the system settles again in virial 
equilibrium and
obeys to the $I^o$ Principle yields the equipartition of the Clausius' virial energy variation 
that is one half of it has to be exchanged with the Univese,
the other half has to contribute to change the 
"temperature"  of the system  which is proportional to the total kinetic energy. 
The result is formally the same found for the single component for which the Clausius' virial energy
coincides with the self potential energy, but the difference for the $B$ component 
inside the $D$ one
is that the potential energy $V_B$ now is a non-monotonic function of $x$ (Fig.1).
The consequence is that the differential eq.(\ref{varvir1}) becomes
the following:
\begin{equation}
\label{DQ}
\Delta\widetilde{Q}=\frac12\Big(\frac{\nu_{\Omega B}}{x^2}-\nu'_Vm(2-d)x^{1-d}\Big)\Delta x
\end{equation}
where $\Delta \widetilde{Q}$ is the exchanged amount of heat 
in normalized units (see, eq.\ref{virn}) and $\Delta x$ the normalized variation
of the $B$ structure semimajor axis.
Moreover, by defining the entropy variation of the $B$ system, in arbitrary units, as:
\begin{equation}
\label{DS}
\Delta\widetilde{S}=\frac{1}{\widetilde{T_B}} \Delta\widetilde{Q}
\end{equation}
The eqs.(\ref{DS}) and (\ref{virn}) yield:
\begin{equation}
\label{DS1}
\Delta\widetilde{S}=
\Big(\frac{\frac{\nu_{\Omega B}}{x^2}-\nu'_Vm(2-d)x^{1-d}}{\frac{\nu_{\Omega B}}{x}+\nu'_Vmx^{2-d}}
\Big)\Delta x
\end{equation}

where $\widetilde{T_B}=-\frac12\widetilde{V_B}$, according to eq.(\ref{fun}).
It is easy to see that the derivative of heat is stationary at $x=x_t$, and as consequence 
also the derivative of the $B$ entropy. Then we have to wait a minimum or a maximum
of the entropy of the luminous component at the {\it tidal radius configuration}.
Indeed, the integration
of eq.(\ref{DS1}) yields always a non-monotonic trend for the function $\widetilde{S}(x)$ with a 
maximun at the 
{\it tidal radius}, as soon as the the maximun in Clausius' virial energy does exist.
In the special case of Tab.1, $b=0;d=0.5$, we obtain:
\begin{equation}
\label{fentro}
\widetilde{S}(x)-\widetilde {S}(1)=ln(12539)-ln\Big(\frac{11339 ~x^{5/2}+1200}{x}\Big)
\end{equation}
Its trend is shown in Fig.3.\\
The existence of this entropy maximum is able to stress the relevance of the Clausius' energy maximum.
Indeed, according to Layzer (1976), the definition of thermodynamic information is :
\begin{equation}
\label{inf}
I=S_{max}-S
\end{equation}
where $S_{max}$ means the maximum value the entropy of the system may have as soon as the constraints
on it, which fix the actual value of its entropy to $S$, are relaxed. Then, in order to 
increase its 
information a system has to decrease its entropy in respect to that of the universe. For a luminous 
component as $B$, which is embedded in an other $D$, a contraction decreases the entropy of 
$B$ only if its
dimension is smaller or equal to $a_t$. It means that only from this dimension downwards the 
component $B$ may gain information during its evolution due to dissipation until to become more 
structured by changing its gas amount into stars (Secco 1999).
Even if the argument has to make wider and deeper in a future work it seems to us it may 
become the ground
for the interpretation of the oserved cut-off radius  observed in many edge-on spiral galaxies
(see, e.g., van der Kruit, 1979;
Pohlen et al. 2000a, 2000b and Kregel et al., 2002). From a preliminary
analysis the observed cut-off adius correlates 
with the $B$ mass as the 
tidal radius does
(Guarise et al. 2001; Secco \& Guarise 2001). 

\begin{figure}[!h]
\begin{center}
\includegraphics[height=9cm,width=12cm]{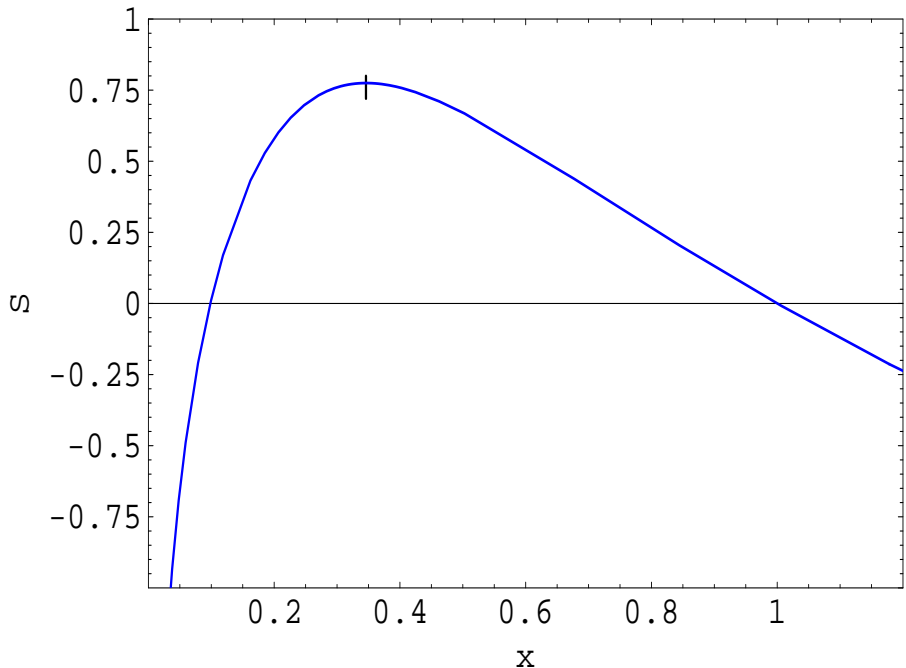}
\caption{Entropy trend of the $B$-system, in arbitrary units, as 
function of $x$ (case 2 of Tab.1). The normalized factor is the same of the other figures. The maximum occurs
at the {\it tidal radius} $x_t=0.346$.}
\label{marmo3}
\end{center}
\end{figure}

\section{DM energy trends}
\label{DM}
By looking at Fig.4 which shows the energy trends of the $D$-component, the 
non-specularity  of the virial energies behavior between the two 
components is manifest.  Indeed, $\widetilde{V}_D$ is
monotonic (curve a), moreover $(\widetilde{E}_{pot})_D$ increases very slowly at decreasing $x$.

It should be underlined that the two energies become equal as soon as the residual energies,
given by the eq.(\ref{qnorm}), are zero.
That occurs at the configuration characterized by the size ratio:
\begin{equation}
\label{zeroq}
x_o\simeq \Big [ \frac{(3-d)}{(3-b)}~\frac{[5- (b+d)]}{(5-2d)}\Big]^{\frac{1}{2-d}}
\end{equation}
If the mass distributions of the two components are the same ($b=d$), the equality appears
only when the two configurations are coinciding ($x_o=1$). In the cases of different mass 
distributions, it may be
possible to find that the virial energy becomes equal to the potential energy only at one
configuration which satisfies the  
physical condition: $0<x<1$; in turn, that means :$$(2-d)(b-d)<0$$ i.e.,
$b<d$ (the case of the Fig.4), if we have taken into account the dynamics of two-component system
at the presence of a 
tidal radius ($d<2$). Then, the difference between the energies, at $x=1$, becomes:
\begin{equation}
\label{score}
(\widetilde {Q}_{BD})_{x=1}=-(\widetilde {Q}_{DB})_{x=1}\simeq \frac{1}{2}\nu'_V~m\frac{(b-d)}{(3-b)}
\end{equation}
All that is true not only for the $D$-component but, of course, also for the $B$-one, due to 
the anti-symmetric
nature of the residual terms.

\begin{figure}[!h]
\begin{center}
\includegraphics[height=9cm,width=12cm]{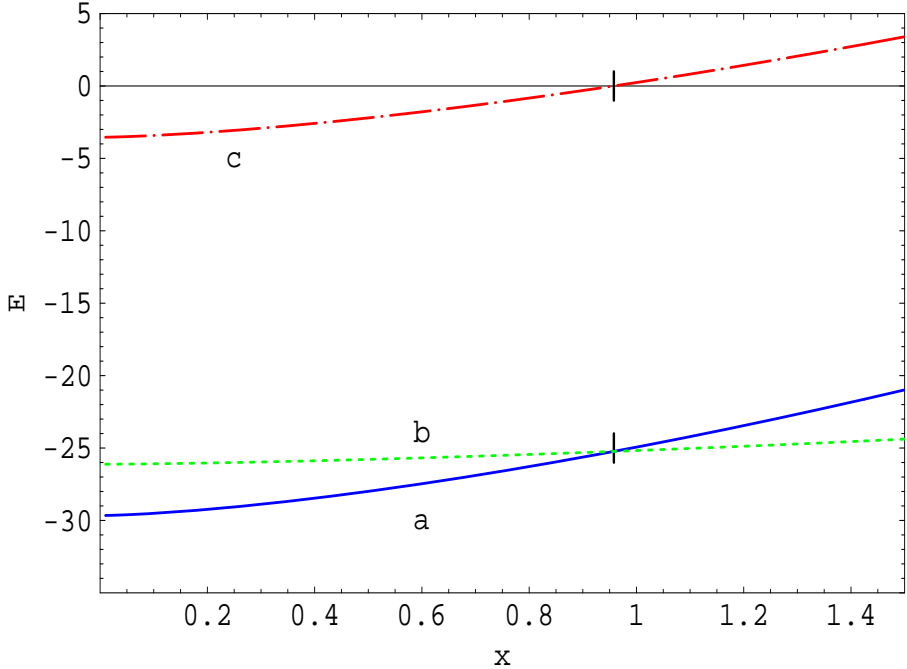}
\caption{The trends of the $D$-system energies, as 
function of $x$ (case 2) of Tab.1), normalized as in the previous figures. The a-curve 
(solid line) and the b-one
(dotted line) represent the behavior of the Clausius' virial, 
$\widetilde {V}_D$,  and of the total potential energy $(\widetilde {E}_{pot})_D$, respectively ; 
the c-curve 
(dash dotted -line) shows the trend
of the residual energy $\widetilde {Q}_{DB}=-\widetilde {Q}_{BD}$. The two energies are
equal at $x_o= 0.9579$ (the vertical mark) and the 
score at $x=1$, is: $(\widetilde {Q}_{DB})_{x=1}= 0.2361$, eq.(\ref{score}).}
\label{marmo4}
\end{center}
\end{figure}

\section{The energies of the whole system}

In the last step we will investigate the potential energy trend of the whole system, which means
the trend of the function (\ref{eptnor}) at decreasing $x$. It is shown in Fig.5.
Again a flex point appears.
Indeed, the first and second derivative turn out to be, respectively:
\begin{eqnarray}
\label{first1}
\frac{d\widetilde {E}_{
pt}}{dx}\simeq \frac{\nu_{\Omega_B}}{x^2}
+\nu'_Vm  
 x^{1-d} \\
\label{second1}
\frac{d^2\widetilde {E}_{
pt}}{dx^2}\simeq -2\frac{\nu_{\Omega_B}}{x^3}
+\nu'_Vm  
 \frac{1}{x^{d}}(1-d)
\end{eqnarray}

\begin{figure}[!h]
\begin{center}
\includegraphics[height=9cm,width=12cm]{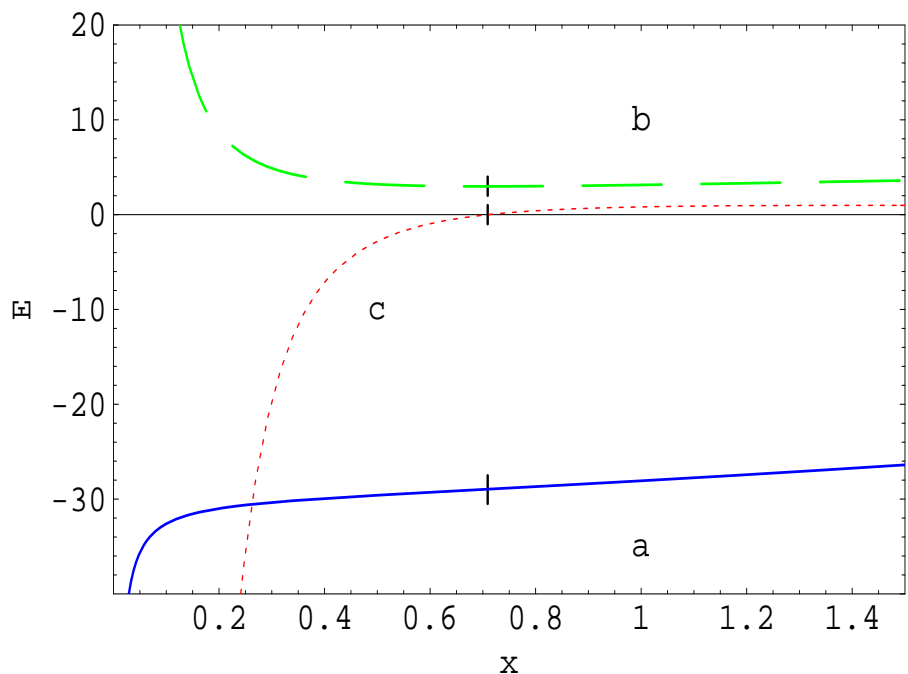}
\caption{The potential energy trend of the whole ($B+ D$)-system, normalized to the usual factor. 
The a-curve (solid line) represents 
the total potential energy, $\widetilde {E}_{pt}$; the b (dashed line) and c-curve (dotted line)
represent the first and second derivative of the total potential energy, respectively.
The curve parameters are those of the case 2) in Tab.1. The flex point of eq.(\ref{tflex}) is marked.}
\label{marmo5}
\end{center}
\end{figure}

Then the flex point will occur for:
\begin{equation}
\label{tflex}
x'_{f}\simeq \big(2\frac{2-d}{1-d}\big)^{\frac{1}{3-d}}x_t=x_{f}/2^{1/(3-d)}
\end{equation}

It should be noted that again the flex point may exist only if $d<1$ and it
is located a factor $1/2^{\frac{1}{3-d}}$ before the flex point
of $(\widetilde{E}_{pot})_B$.

The total potential energy trend
of Fig.5 for the whole system is also that of its total mechanical energy when it is in virial 
equilibrium. Indeed,
according to eq.(\ref{eq7}), we have:
\begin{equation}
\label{mec}
E_{mec}= E_{pt}+T_B+T_D=V_B+V_D-\frac{1}{2}V_B--\frac{1}{2}V_D=\frac{1}{2}E_{pt}
\end{equation}
The result is different for the single u-subsystem. From eqs.(\ref{eq5},\ref{eq6}) we obtain:
\begin{equation}
\label{mecu}
(E_{mec})_u= (E_{pot})_u+T_u=\Omega_u+W_u-\frac{1}{2}V_u=
\frac{1}{2}V_u-Q_{uv}
\end{equation}
The trends of the u-mechanical energies are given in Fig.6. 
It is difficult to look for a physical meaning related to the small increase of 
the mechanical energy of the $D$ component as $x$ decreases (c curve). It is probably due to the
artificial constraint, we have considered, that the dark matter system has a size and a mass distribution
insensitive to the 
contraction of the baryonic component which occurs inside. As already pointed out by Barnes \& White
(1984), there is an iduced effect from $B$ density distribution on the inner regions of dark halo which,
even if non-dramatic, might change the coefficient $\nu_{\Omega D}, (\nu_{D})_M$ and then $\nu'_V$.
 
\begin{figure}[!h]
\begin{center}
\includegraphics[height=9cm,width=12cm]{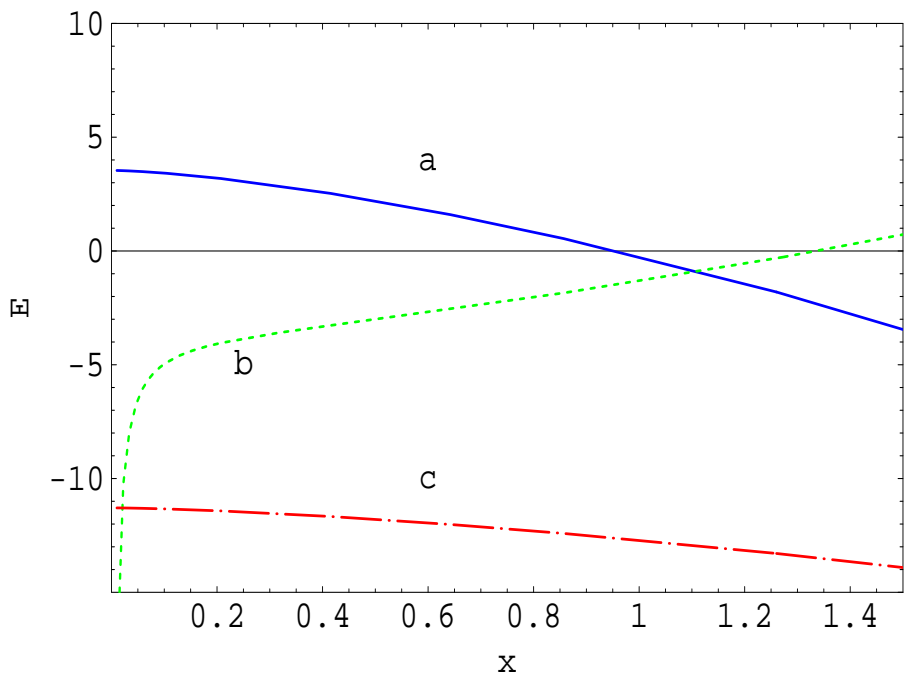}
\caption{The mechanical energy trends of the u-subsystem, given by the 
eq.(\ref{mecu}), normalized 
to the usual factor.
The b (dotted line), c-curve (dash dotted-line) and a-curve (solid)
represent the $(\widetilde {E}_{mec})_{B}$, $(\widetilde {E}_{mec})_{D}$ and the 
residual energy $\widetilde{W}_{BD}$, respectively.
The curve parameters are those of the case 2) in Tab.1.}
\label{marmo6}
\end{center}
\end{figure}

\section{Discussion and Conclusion}

The energetics of a two-component virial system has been analized during its dynamic evolution due to
the contraction of a inner (baryonic) component inside an outer of dark matter, which is 
assumed, for
sake of simplicity, to be fixed. The results are also compared with those we obtain without
the presence of the dark potential well. The analysis was done with the help of some simple,
good approximate expressions for all the energies (in the cases of $b<2$ and with $\xi_c=0.01$) which, 
once the form factor is assigned, are only 
functions of the $b$ and $d$
exponents in the two power-law density distributions.
The different character between the energy trends of Clausius' virial and of the total potential energy for
the baryonic (or bright component $B$), has been underlined and the physical main 
consequences has been derived
for the minimum of the virial energy.
The particular location of the special $B$ size corresponding to the minimum ({\it tidal radius}) on 
the total potential energy curve and the discover of one flex point on it, has stressed 
the character of the {\it tidal radius} as a scale length induced by the dark
matter halo component on the gravitational field of the inner baryonic one. 
This scale length marks the beginning
of the two different regimes as follows: the self-energy regime, when $x<x_t$, and  
the tidal-energy regime at about $3x_t$.

We claim as observable counter-parties of these theoretical reasons 
for the real existence of this induced scale length, at least the two followings: the first one is 
more direct:
it is the cut-off radius observed in many edge-on spiral galaxies. Thermodynamical 
arguments related to the presence of an entropy maximum at this scale length and
some preliminary comparison between correlations from observables and theoretical expectations
seem to prove the connection.
The other one is more indirect but it has an important, observable consequence:  
if there is this length, we are able to 
reproduce some of the main
features of the Fundamental Plane (FP) for the elliptical galaxies (LS1).
From a general point of view, the two arguments are connected by the 
thermodynamics in the following way: the elliptical galaxies stay on the 
maximum entropy stage without a significant structure evolution, the cut-off spirals,
practically, begins their evolution from this stage forwards.

How the main physical results depend on the approximations used has also been considered.
The main result is that: a too small homogeneous cores in both the two subsystems might cancel
this induced length as soon as $b\ge2.5$.
Avoiding the divergency at the center of the two configurations, seems
to become an essential condition in order to obtain the FP features. Here we are constrained by the
restriction of handling together the two cores and then we need of a further paper on this matter in order to 
disentangle the probable different core-dimension play.

Nevertheless, from this point of view an appealing application to the galaxy clusters may be
thought of. If we take into account the values of the exponent $\alpha$, used by Girardi et al.(1995) 
in the power-laws
which give the best fit for the galaxy distributions, at a distance $r>> R_c$ ($R_c$= core radius),
in the clusters, we can see that the corresponding range of our exponent $b$
turns out to be approximately the critical one 
we have found in the section (\ref{BM}), $2.5\le b ~<~3$, with ratios $R_c/R_{vir}$ (our $\xi_c$) which are big enough for
obtaining the maximum virial energy at $x<1$.

The energies of the $D$-subsystem were also considered. Its total potential energy is practically
unaffected by the contraction of the $B$-subsystem whereas the virial energy
slowly increases,
in absolute value, as $x$ decreases. The conditions to obtain the 
equality between these two energies was 
with the main result that only one configuration may enjoy this property. It corresponds to the
initial one, when the two substructures fill the same volume, only if the two mass distributions
are the same.

Finally the trend of the potential energy of the whole system tells us that a configuration exists
of minimum variation of the total potential energy (and then of the total mechanical energy), a flex 
point,
inside its monotonic behavior. The conditions required in order to obtain either the flex on the $B$
total potential energy , or on the total potential energy of the whole system 
were investigated
and the result again is a constrain only on the dark matter distribution (see the {\it tilt} of the FP).
The mechanical energies either of the whole system or
of both the subsystems were also taken into account.

 Even if we are aware that the next step we have to do is to investigate the 
energetics of a two-component system with or without central divergencies and
characterized by two more realistic density distributions, e.g., of 
Hernquist (1990) kind for the inner component and of  Zhao (1996) general kind for the dark 
matter halo (included the NFW profile \citep {nav}), as already considered, e.g., by 
Caimmi \& Marmo (2002),
the present results seem to be very attractive for the future work.

\section*{Acknowledgements}

 We thank our friend, Prof. R. Caimmi for helpful discussions
 and mathematical support.

\appendix
\section{Appendix}
According to RC we define:
\begin {eqnarray}
\label{wint}
w^{int}(x)= L(\xi_c,x) + M(\xi_c,x)\\
\label{wex}
w^{ext}(x)= G(\xi_c,x) + H(\xi_c,x)\\
\label{fli}
L(\xi_c,x)=\int_{0}^{\xi_c}F_D\frac{dF_B}{d\xi_D}
\xi_D~d\xi_D \\
\label{fmi}
M(\xi_c,x)=\int_{\xi_c}^{x }
F_D\frac{dF_B}
{d\xi_D}\xi_D~d\xi_D\\ 
\label{fgi}
G(\xi_c,x)=\int_{0}^{\xi_c}F_B\frac{dF_D}
{d\xi_D}\xi_D~d\xi_D \\
\label{acca}
H(\xi_c,x)=\int_{\xi_c}^{x }
F_B\frac{dF_D}
{d\xi_D}\xi_D~d\xi_D 
\end{eqnarray}
The two functions $F_B(\xi_B=\frac{a_D}{a_B}\xi_D)$ and $F_D(\xi_D)$ in the ranges $(0,\xi_c)$ and
$(\xi_c,x)$, for different values of 
exponents $b$ and $d$, are given in Tab.3 of Appendix A in LS, with the assumption that 
$(\xi_c < \frac{a_B}{a_D})$. 

In Table 1 of Appendix A (LS1), the corresponding functions
$L(\xi_c,x)$ and $M(\xi_c,x)$ are shown
with the approximate analytical expressions of $w^{int}(x)$ 
we will use in the next sect.. In Table 4 of Appendix A (LS), the corresponding functions
$G(\xi_c,x)$ and $H(\xi_c,x)$ are given
with the approximate analytical expressions of $w^{ext}(x)$ 
which we will also need.
 We take into account the approximation
of $w^{ext}$, 
towards which all the exact analytic expressions of it converge, by disregarding terms of higher
order in respect to $\xi_c$ at a different degree (see Table 4 in Appendix A of LS), which is:
\begin{equation}
\label{wext}
w^{ext}(\frac{a_B}{a_D})\simeq w^{ext}_{prox}=-\frac{4\xi_c^{b+d}}{(3-d)[5-(b+d)]}(\frac{a_B}{a_D})^{3-d}
\end{equation}
despite the $b$ and $d$ values, provided the
following limitations hold:
\begin{equation}
\label{limits}
0\le b <~3~;~ 0\le d <~2 \Rightarrow (b+d)< 5
\end{equation}
These limitations also ensure that the tidal
and the self potential-energy tensor have the same sign, in 
all the cases we take into account.
In order to highlight the $D$ mass fraction, $\widetilde {M}_D=M_D(\frac{a_B}{a_D})^{3-d}$, which 
exerts
dynamic effects on $B$ according to Newton's theorem (the equivalent of $M^*_D$ defined 
for the homogeneous case, see LS), we define the factor:
\begin{equation}
\label{nupv}
\nu'_V=\frac{9}{2}[(\nu_B)_M(\nu_D)_M]^{-1} 
 \frac{\xi_c^{b+d}}{(3-d)[5-(b+d)]}
\end{equation}
In this way it follows that:
\begin{equation} 
\label{Anuv}
\nu_V\simeq \nu'_V \frac{M_D}{M_B}(\frac{a_B}{a_D})^{3-d}
\end{equation}
Moreover under the limitations given by the (\ref{limits}), the following 
approximation also holds (see LS1):
\begin{eqnarray}
\label{wintt}
w^{int}(\frac{a_B}{a_D})\simeq \frac{4\xi_c^{b+d}}{(2-d)[5-(b+d)]}
(\frac{a_B}{a_D})^{3-d}-\frac{4\xi_c^{b+d}}{(2-d)(3-b)}(\frac{a_B}{a_D}) + \epsilon\\
\epsilon=-4\frac{\xi_c^{d+3}}{(2-d)3}(\frac{a_D}{a_B})^2\delta\\
\delta=\Big[1-\frac{3}{3-b}\Big(\frac{a_B}{a_D}\Big)^b\Big]
\end{eqnarray}

As we have already underlined (LS1) in case $b=0$, $\delta$ and 
then $\epsilon$ of eq.(\ref{wintt}) become $0$ , whatever 
the value of $d$ may be,
and that the relevance of $\epsilon$, in respect to the other terms of the 
same equation, increases quickly,
at fixed $d$, as soon as $b$ increases. Moreover, at fixed $b$, the weight 
of $\epsilon$ 
grows as $d$ increases,
even if
less than before; at fixed $b$ and $d$ the contribution of $\epsilon$ increases as
the ratio $a_B/a_D$ becomes smaller. If we limit 
ourselves to choosing $b\leq1.5$ and $d\simeq0.5$, we may disregard 
$\epsilon$ in the approximation of $w^{int}$ with a maximum error of $12\%$ 
when $b=1.5$ and $a_B/a_D$
decreases to $0.4$. That occurs if we choose $\xi_c=0.1$, but the error decreases as $\xi_c$ decreases. 

It should be noted that 
the approximation of $w^{int}$ is different from that of $w^{ext}$, even if 
the core dimensions $\xi_c$, 
are the same, because of the
$b$ value. Indeed, for $2\appleq  b$ one needs to introduce in the approximated $w^{int}$ the term 
$\delta$ in which the ratio $a_B/a_D$
has $b$ as exponent. That does not appear in the $w^{ext}$ approximation.

By also taking into account the approximation of eq.(\ref{wext}) the
following relation holds, as soon as $\epsilon$ is negligible:
\begin{equation}
\label{Approx}
w^{int}\simeq -\frac{4\xi_c^{b+d}}{(2-d)(3-b)}(\frac{a_B}{a_D})-w^{ext}_{prox}\frac{(3-d)}{(2-d)}
\end{equation}

By using eq.(\ref{Approx}), ($\epsilon \simeq 0$), we obtain the following 
approximation for the coefficient of eq.(\ref{nuvd}), as a function of $\nu'_V$:
\begin{equation}
\label{nuaprx}
\nu_{VD}\simeq\nu'_V\frac{1}{m}\frac{3-d}{2-d}\Big\{\frac{5-(b+d)}{3-b}-(\frac{a_B}
{a_D})^{2-d}\Big\}
\end{equation}

\end{document}